\documentclass[preprint,superscriptaddress,prb]{revtex4}

\usepackage{graphicx}

\begin{document}

\title{Ultra-long wavelength Dirac plasmons in graphene capacitors}

\author{H. Graef} \affiliation{Laboratoire Pierre Aigrain, Ecole normale sup\'erieure, PSL University, Sorbonne Universit\'e, Universit\'e Paris Diderot, Sorbonne Paris Cit\'e, CNRS, 24 rue Lhomond, 75005 Paris France}
\author{D. Mele} \affiliation{Laboratoire Pierre Aigrain, Ecole normale sup\'erieure, PSL University, Sorbonne Universit\'e, Universit\'e Paris Diderot, Sorbonne Paris Cit\'e, CNRS, 24 rue Lhomond, 75005 Paris France}
\author{M. Rosticher} \affiliation{Laboratoire Pierre Aigrain, Ecole normale sup\'erieure, PSL University, Sorbonne Universit\'e, Universit\'e Paris Diderot, Sorbonne Paris Cit\'e, CNRS, 24 rue Lhomond, 75005 Paris France}
\author{C. Stampfer}
\affiliation{JARA-FIT and 2nd Institute of Physics, RWTH Aachen University, 52056 Aachen, Germany}
\author{T. Taniguchi}
\affiliation{Advanced Materials Laboratory, National Institute for Materials Science, Tsukuba,
Ibaraki 305-0047,  Japan}
\author{K. Watanabe}
\affiliation{Advanced Materials Laboratory, National Institute for Materials Science, Tsukuba,
Ibaraki 305-0047, Japan}
\author{E. Bocquillon}\affiliation{Laboratoire Pierre Aigrain, Ecole normale sup\'erieure, PSL University, Sorbonne Universit\'e, Universit\'e Paris Diderot, Sorbonne Paris Cit\'e, CNRS, 24 rue Lhomond, 75005 Paris France}
\author{G. F\`eve}\affiliation{Laboratoire Pierre Aigrain, Ecole normale sup\'erieure, PSL University, Sorbonne Universit\'e, Universit\'e Paris Diderot, Sorbonne Paris Cit\'e, CNRS, 24 rue Lhomond, 75005 Paris France}
\author{J-M. Berroir}\affiliation{Laboratoire Pierre Aigrain, Ecole normale sup\'erieure, PSL University, Sorbonne Universit\'e, Universit\'e Paris Diderot, Sorbonne Paris Cit\'e, CNRS, 24 rue Lhomond, 75005 Paris France}
\author{E.H.T. Teo}
\affiliation{School of Electrical and Electronic Engineering, Nanyang Technological University, Singapore}
\author{B. Pla\c{c}ais} \email{bernard.placais@lpa.ens.fr}
\affiliation{Laboratoire Pierre Aigrain, Ecole normale sup\'erieure, PSL University, Sorbonne Universit\'e, Universit\'e Paris Diderot, Sorbonne Paris Cit\'e, CNRS, 24 rue Lhomond, 75005 Paris France}

\begin{abstract}
Graphene is a valuable 2D platform for plasmonics as illustrated in recent THz and mid-infrared optics experiments. These high-energy plasmons however, couple to the dielectric surface modes giving rise to hybrid plasmon-polariton excitations. Ultra-long-wavelengthes address the low energy end of the plasmon spectrum, in the GHz-THz electronic domain, where intrinsic graphene Dirac plasmons are essentially decoupled from their environment. However experiments are elusive due to the damping by ohmic losses at low frequencies. We demonstrate here a plasma resonance capacitor (PRC) using hexagonal boron-nitride (hBN) encapsulated graphene at cryogenic temperatures in the near ballistic regime. We report on a $100\;\mathrm{\mu m}$ quarter-wave plasmon mode, at $40\;\mathrm{GHz}$, with a quality factor $Q\simeq2$. The accuracy of the resonant technique yields a precise determination of the electronic compressibility and kinetic inductance, allowing to assess residual deviations from intrinsic Dirac plasmonics. Our capacitor GHz experiment constitutes a first step toward the demonstration of plasma resonance transistors for microwave detection in the sub-THz domain for wireless communications and sensing. It also paves the way to the realization of doping modulated superlattices where plasmon propagation is controlled by Klein tunneling.
\end{abstract}

\date{today}

\maketitle

\begin{figure}[hh]
\centerline{\includegraphics[width=16cm]{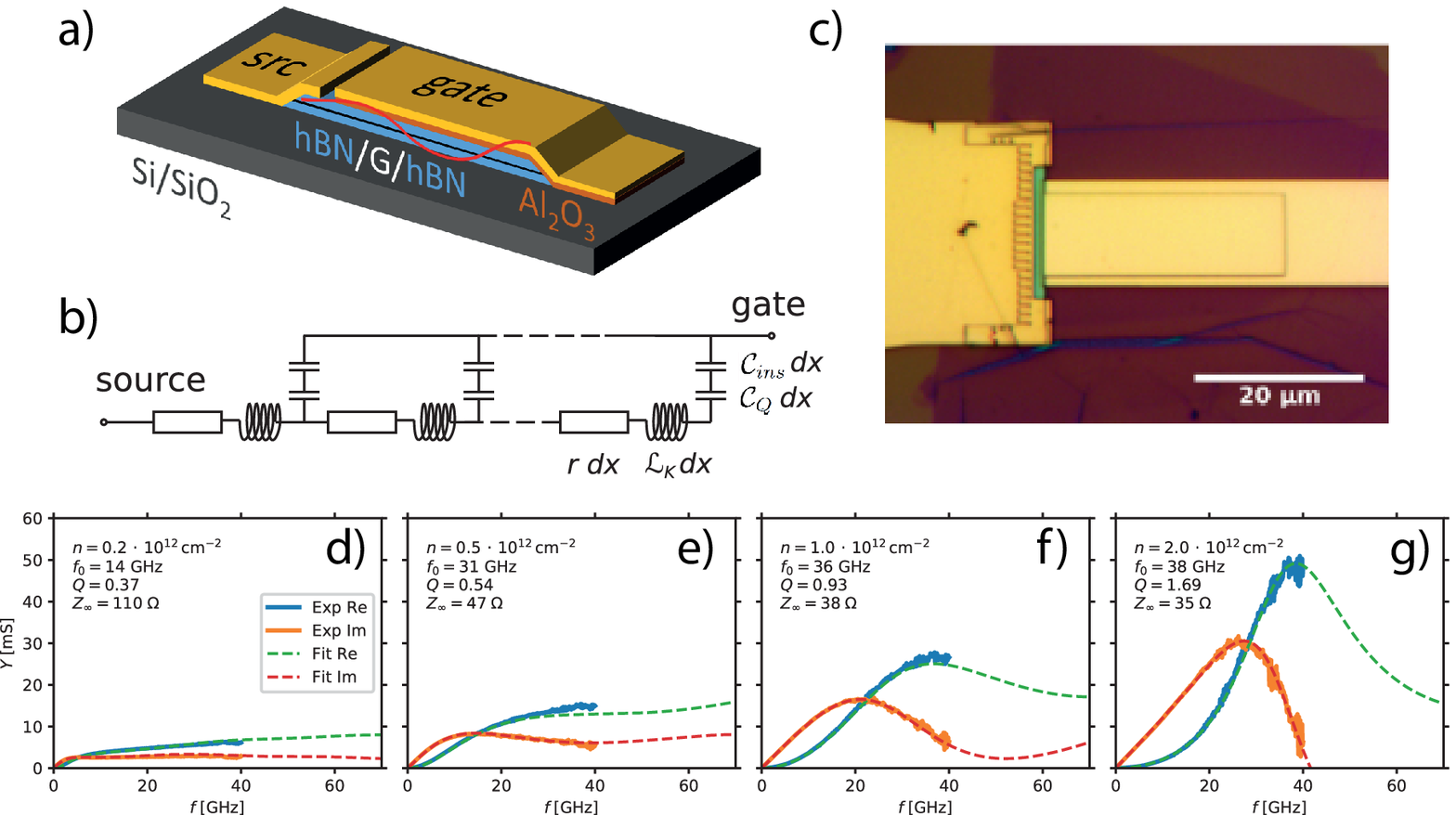}}
\caption{Admittance spectrum of a graphene plasma resonance capacitor (PRC). a) sketch of a PRC made of encapsulated graphene (blue/black) equipped with and edge contact and a top gate (golden). A thin layer of Al$_2$O$_3$ (brown) is used to isolate graphene from the top gate. b) distributed line model of the graphene resonator. c) optical image of a typical T-shape encapsulated PRC sample obtained by etching a broad BN-graphene-BN stack.
Panels d-g) complex admittance of PRC-D6  (dimensions $L\times W=24\times8\;\mathrm{\mu m}$) measured at $T=30\;\mathrm{K}$ in increasing electron density $n\simeq 0.2$--$2\;10^{12}\;\mathrm{cm^{-2}}$. For clarity the theoretical fits with Eq.(\ref{equation}) (dashed lines) are shown in an extended frequency range exceeding the $40\;\mathrm{GHz}$ bandwith of our measuring probe station. The carrier density and fitted values of the resonant frequency $f_0$, quality factor $Q$, and characteristic impedances $Z_\infty$ are specified in the figures. The plasmon resonance is overdamped ($Q\simeq0.4$) at $n\simeq 0.2\;10^{12}\;\mathrm{cm^{-2}}$ in panel d) due to increased ohmic losses. It is well developed at $n= 2\;10^{12}\;\mathrm{cm^{-2}}$ in panel g) with $Q=1.69$ at $f_0=38\;\mathrm{GHz}$. From the admittance maximum   $\Re(Y)(f_0)=2/R=50\;\mathrm{mS}$ we deduce a channel resistance $R=40\;\mathrm{Ohms}$.}
 \label{PRC-principles.fig1}
\end{figure}

Two dimensional electron systems (2DES) sustain both single-particle and collective low-energy excitations, the latter being called plasmons. In graphene their interplay is controlled  by the electron density $n$ which rules the kinetic energy, interactions and damping. Free 2DES plasmons are dispersive with $\omega_p\propto\sqrt{q}$ and a velocity $v_p$ diverging in the long wavelength limit [\onlinecite{DasSarma2011rmp}]. However, plasmons are screened in the long wavelength limit $\lambda_p\gtrsim d$ [\onlinecite{Burke2000apl}], where $d$ is the gate-2DES distance, and acquire an energy-independent velocity $v_p=v_F\sqrt{(e^2d/\pi\epsilon)(k_F/\hbar v_F)}$, where $v_F$ and  $k_F=\sqrt{\pi n}$ are the Fermi velocity and wave vector, and $\epsilon\simeq3\epsilon_0$ the hBN permittivity. Plasmons have been mostly investigated in the THz and mid-infrared (MIR) optics domains where the damping length $\alpha^{-1}\gtrsim10\;\mathrm{\mu m}$ was found to widely exceed the wavelength $\lambda_p\lesssim1\;\mathrm{\mu m}$ [\onlinecite{Knap2002apl,Grikorenko2012nphoton,Lundeberg2017science,Ni2018nature}]. In this high-energy range, plasmons couple to the dielectric surface modes forming plasmon-polaritons states [\onlinecite{Ni2018nature}].  Ultra-long-wavelength plasmons ($\lambda_p\sim100\;\mathrm{\mu m}$) belong to the GHz domain and  do not suffer from this hybridization. They have been observed in high-mobility 2DESs in  Ref.[\onlinecite{Burke2000apl}], but remain elusive in graphene in spite of the interest of manipulating collective chiral Dirac fermion excitations. A constitutive element of plasmon propagation is the kinetic inductance which has been measured in Ref.[\onlinecite{Yoon2014nnano}]. High-mobility graphene [\onlinecite{Banszerus2016nl}] offers the opportunity to investigate GHz plasmonics with a transport approach. Motivations are manyfold; they include the demonstration of GHz plasma resonance devices [\onlinecite{Ryzhii2012jpap}], the investigation of plasmonic crystals in doping-modulated structures [\onlinecite{Aizin2012prb}], that of interactions including viscous fluid effects at high temperature [\onlinecite{Bandurin2016science,Crossno2016science}], or the coupling with hyperbolic phonon polaritons of the hBN dielectric [\onlinecite{Ni2018nature,Tielrooij2018nnano,Yang2018nnano}] at high bias.
In this letter we demonstrate a plasma resonance capacitor (PRC in Fig.\ref{PRC-principles.fig1}-a) where $100$-$\mathrm{\mu m}$ Dirac plasmons propagate along a hBN-graphene-hBN-metal strip line of length $L\simeq24\;\mathrm{\mu m}$ and aspect ratio $L/W=3$ (Fig.\ref{PRC-principles.fig1}-c). We report on the fundamental mode, which is a quarter-wave resonance at $f_0=v_p/4L\sim40\;\mathrm{GHz}$, where we achieve a quality factor  $Q\simeq2$ at low temperature ($T=10\;\mathrm{K}$) corresponding to a damping length $\alpha^{-1}=2QL/\pi\sim30\;\mathrm{\mu m}$.

Our graphene resonator is a T-shape hBN-encapsulated single-layer graphene sample covered by a top metallization serving both as radio-frequency (RF) port and DC gate (see methods, Fig.\ref{PRC-principles.fig1}-c and Figs.\ref{PRC_techno.fig3}). We have measured a series of eight devices, using both exfoliated and CVD graphene, but results presented below focus on sample PRC6 ($L\times W=24\times 8\;\mathrm{\mu m}$) which has the highest mobility and largest quality factors. The device is embedded in a coplanar waveguide and its RF gate-source admittance $Y(f,n,T)$ is measured in the range $f=0$--$40\;\mathrm{GHz}$, $n=0$--$2\times10^{12}\;\mathrm{cm^{-2}}$ and $T=10$--$300\;\mathrm{K}$ using standard vector-network-analyzer (VNA) techniques (see methods and Ref.[\onlinecite{Chaste2008nl}]). The strip line access, which has a broader width, constitutes the device source. It is equipped with low-resistance edge contacts (Fig.\ref{PRC-principles.fig1}-c). The source end of the strip is an impedance short, securing a plasmon node, the open end of the strip being an antinode so that the PRC sustains odd harmonics of the fundamental mode $f_0=v_p/4L$.

We describe propagation by a distributed-line model (Fig.\ref{PRC-principles.fig1}-b). The line capacitance $\mathcal{C}$ is the series addition of the insulator capacitance $\mathcal{C}_{ins}=\epsilon W/d$  and the quantum capacitance $\mathcal{C}_Q =\frac{4e^2}{h}\;k_FW/v_F$ (low temperature). The large top-gate capacitance  enhances the  $\mathcal{C}_Q$ contribution and gives access to a capacitance spectroscopy which is used below to determine the Dirac-point position. The line inductance is dominated by the kinetic contribution, which at low temperature writes  [\onlinecite{Yoon2014nnano}]
\begin{equation}
\mathcal{L}_K =\frac{h}{4e^2}\frac{2}{k_Fv_FW}\; .
\label{inductance}
\end{equation}
 From these expressions one recovers the plasmon velocity $v_p=1/\sqrt{\mathcal{L}_K \mathcal{C}}$ and  the characteristic impedance $Z_\infty =\sqrt{\mathcal{L}_K/ \mathcal{C}}$ of the plasmonic strip line. The line resistance  $r=R/L$ accounts for plasmon losses which in principle include ohmic, viscous and dielectric contributions. The latter are negligible in our low-frequency range and viscous losses are minimized by using a plane wave geometry. Upon increasing the carrier density the plasmonic response, with $\mathcal{L}_K\propto1/k_F$ (single layer graphene), takes over single particle diffusive response, with $r\propto1/k_F^2$ (constant mobility), allowing for low-damping plasmon propagation. The crossover occurs whenever the total strip resistance $R\lesssim Z_\infty$, or equivalently the plasmon damping length $\alpha^{-1}\gtrsim L$. Dealing with a resonant device, the PRC admittance,  $Y=j \mathcal{C}\omega\times\tanh{\left[L\sqrt{j \mathcal{C}\omega(r+j \mathcal{L}_K\omega)}\right]}/\sqrt{j \mathcal{C}\omega(r+j \mathcal{L}_K\omega)}$ [\onlinecite{Pallecchi2011prb}], is conveniently cast into the compact form :
\begin{equation}
 Y=Z_\infty^{-1}\times\frac{\tan{\left(x\sqrt{1-2j /Qx}\right)}}{\sqrt{1-2j /Qx}}\;,
\label{equation}\end{equation}
where  $x=\omega L\sqrt{\mathcal{C}\mathcal{L}_K}$ is the reduced frequency. The capacitive character of the resonator is encoded in the low-frequency response  $Y/L=j\mathcal{C}\omega$. Eq.(\ref{equation}) features a resonance behavior with a fundamental frequency at $x_0=\pi/2$, an admittance peak amplitude $\Re(Y)(f_0)=Q/Z_\infty=2/R$ and a width $\Delta f \simeq f_0/Q$. It includes higher harmonics, constituting the frequency comb $f_k=(2k+1)f_0$ with $Q_k\sim(2k+1)Q$. The latter are discarded here due to the finite $40\;\mathrm{GHz}$ bandwidth of our cryogenic setup.

\begin{figure}[hh]
\centerline{\includegraphics[width=15cm]{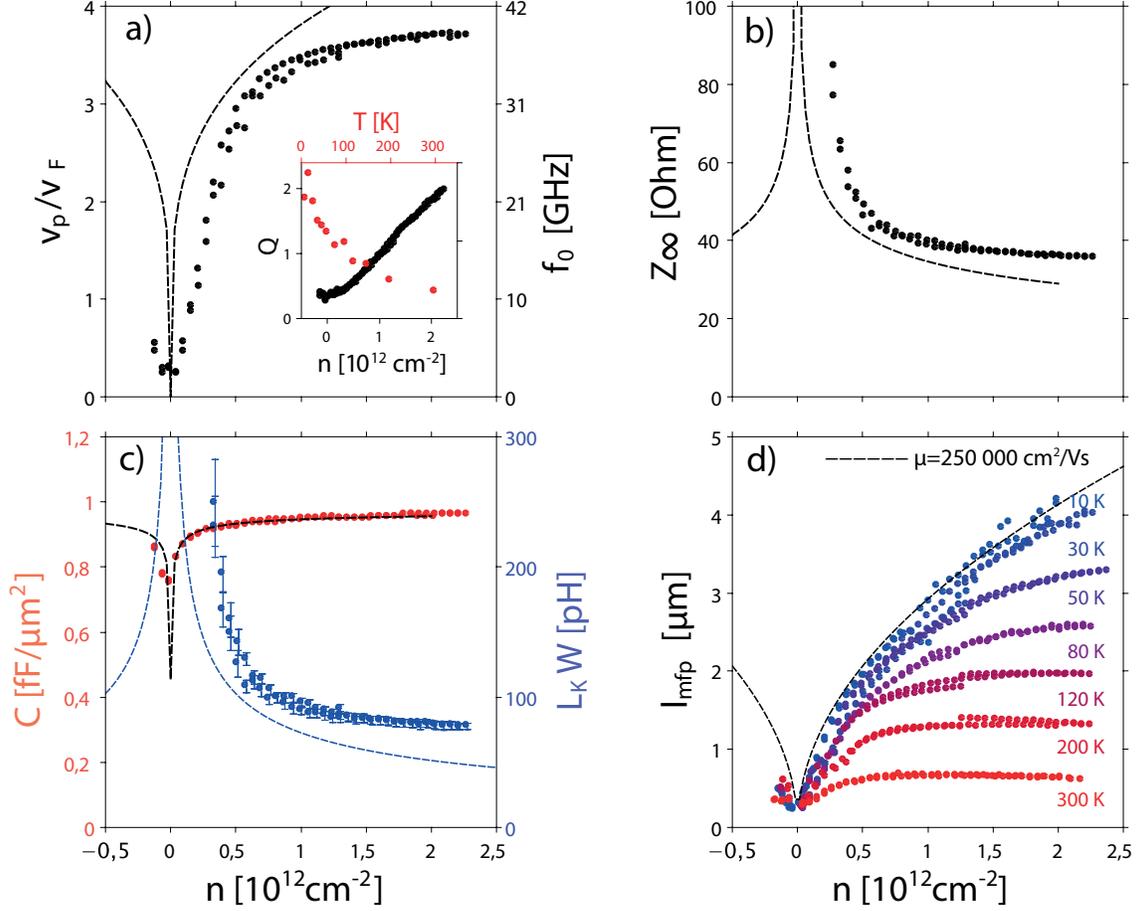}}
\caption{Density and temperature dependence of the graphene plasma resonance capacitor (PRC) deduced from a three-parameter theoretical fits of the admittance spectra with Eq.(\ref{equation}). a) density dependence of the plasmon velocity at $T=30\;\mathrm{K}$, deduced from the resonant frequency $v_p=4Lf_0$. The dashed line is the theoretical expectation of Eq.(\ref{inductance}) for the low temperature limit. Inset, quality factor :  density dependence at $30\;\mathrm{K}$ (black), temperature dependence at $2\;10^{12}\;\mathrm{cm^2/Vs}$ (red).  b) density dependence of the plasmonic line characteristic impedance ($T=30\;\mathrm{K}$); dashed line is the theoretical prediction. c) density dependence of the capacitance (red dots)  and kinetic inductivity (blue dots) $\mathcal{L}_KW$  ($T=30\;\mathrm{K}$). Black and blue dashed lines are the low-temperature theoretical expectations for the capacitance and inductivity $\mathcal{L}_KW$.  d) density dependence of the electronic mean-free-path $l_{mfp}$  for a representative set of measuring temperatures. $l_{mfp}$ saturates at low temperature to a constant mobility regime $l_{imp}=\mu\hbar k_F/e$ with $\mu\simeq250 000\;\mathrm{cm^{2}V^{-1}s^{-1}}$ (black dashed line). At high temperature, we retrieve the acoustic-phonon-limited scattering length plateau $l_{ph}\sim 0.7\times(300/T)\;\mathrm{\mu m}$.}
\label{plasmon-parameters.fig2}
\end{figure}

The rise of a plasma resonance in increasing electron density is illustrated in Figs.\ref{PRC-principles.fig1}-(d-g). Admittance spectra, measured at $T=30\;\mathrm{K}$, have been obtained after deembedding a contact resistance $R_c=43\;\mathrm{Ohms}$. Such a low value was obtained by back-gating the capacitor access region. Data are presented together with their fit with Eq.(\ref{equation}) (dashed lines). At the lowest density the channel resistance takes over the kinetic inductance. The plasmon resonance $f_0\sim 14\; \mathrm{GHz}$ is overdamped with $Q\sim0.4$ (Fig.\ref{PRC-principles.fig1}-d). The admittance spectrum is reminiscent of the evanescent wave response reported in Refs.[\onlinecite{Pallecchi2011prb,Inhofer2017prb,Inhofer2018prapplied}]. Fingerprints of a resonant behavior  become perceptible for $Q\simeq0.5$ with a shallow  minimum of $\Im(Y)$ at $f\simeq f_0$ (Fig.\ref{PRC-principles.fig1}-e). A genuine resonance develops at $n\gtrsim 1\;10^{12}\;\mathrm{cm^{-2}}$ (Fig.\ref{PRC-principles.fig1}-f) and culminates at $f_0\simeq 38\; \mathrm{GHz}$ with $Q=1.69$ for $n=2\;10^{12}\;\mathrm{cm^{-2}}$ (Fig.\ref{PRC-principles.fig1}-g). We deduce a  plasmon velocity $v_p=4Lf_0=3.6\;10^6\;\mathrm{m/s}$ in agreement with the above estimate based on geometry, and a plasmon damping length $\alpha^{-1}=(2Q/\pi)L\simeq 26\;\mathrm{\mu m}$. Plasmon losses correspond to a channel resistance $R=40\;\mathrm{Ohms}$ (Fig.\ref{PRC-principles.fig1}-g) approaching the characteristic impedance $Z_\infty=35\;\mathrm{Ohms}$, which is a suitable measuring condition. The same procedure has been reproduced by recording 400 complex spectra covering the density and temperature ranges, $n=-0.1$--$2.25\;10^{12}\;\mathrm{cm^{-2}}$  and  $T=10$--$300\;\mathrm{K}$, where we find a quality factor peaking to $Q\gtrsim2$ at $10\;\mathrm{K}$ (inset of Fig.\ref{plasmon-parameters.fig2}-a).
Figs.\ref{plasmon-parameters.fig2}-(a,b) summarize the doping dependence of resonator parameters $v_p(n)$, $Z_\infty(n)$ obtained using the same fitting procedure. From these values we calculate $\mathcal{L}_K(n)$, $\mathcal{C}(n)$ ($30\;\mathrm{K}$-data in Fig.\ref{plasmon-parameters.fig2}-c) and find a good agreement with theoretical capacitance formula (black dashed line), including the dip of $\mathcal{C}_Q(n)$ at neutrality. Still we observe a deviation of $\mathcal{L}_K(n)$ with theory (blue dashed line) which exceeds our experimental uncertainties. This deviation is significant thanks to the increased accuracy of the resonant capacitor technique in the determination of  $\mathcal{C}$ and $\mathcal{L}_K$ when compared to non-resonant techniques [\onlinecite{Yoon2014nnano}]. It is too large ($\gtrsim 100\;\mathrm{pH}$) to be explained by geometrical inductance effects ($\mu_0W\sim 10\;\mathrm{pH}$) or systematic errors in the deembedding procedure ($\pm 20\;\mathrm{pH}$ depending on frequency). From the channel resistance we estimate the conductivity and the mean free path $l_{mfp}(n,T)$ which is plotted in (Fig.\ref{plasmon-parameters.fig2}-d) for typical temperatures. Mobility saturates near $\mu=250 000\;\mathrm{cm^2V^{-1}s^{-1}}$ at low temperature (dashed black line in Fig.\ref{plasmon-parameters.fig2}-d). At high temperature we find a mean free path plateau $l_{mfp}(T)\sim0.7\times 300/T\;\mathrm{\mu m}$, in agreement with theoretical estimates the acoustic phonon limited resistivity [\onlinecite{Hwang2008prb}].

To summarize, we have shown that the plasma resonance capacitor principle works and provides an extensive characterization of equilibrium and transport graphene parameters including the compressibility $\mathcal{C}_Q(n,T)$, the kinetic inductance $\mathcal{L}_K(n,T)$ and  mean-free-path $l_{mfp}(n,T)$. The plasmon velocity matches expectations for the screened case. The weak doping dependencies of the plasmon velocity and characteristic impedance in Figs.\ref{plasmon-parameters.fig2}-a,b reflect theoretical expectations for massless graphene where $v_p, Z_\infty\propto n^{-\frac{1}{4}}$ as opposed to $v_p\propto n^{-\frac{1}{2}}$ for massive 2DESs such as bilayer graphene. The deviation from theory, observed in the inductance, gives rise to a saturation of the plasmon velocity $v_p\lesssim 4 v_F$ and characteristic impedance $Z_\infty \gtrsim 35\;\mathrm{Ohms}$. A tentative explanation for this discrepancy is an additive mass contribution from the dilute 2DES in the back-gated silicon substrate, which loads graphene Dirac plasmons according to its capacitive coupling to graphene electrons, and eventually restricts electron mobility. Such a residual substrate coupling can easily be avoided by substituting the silicon back gate with a metallic bottom gate following Refs.[\onlinecite{Yang2018nnano,Wilmart2016scirep}].

In conclusion, we have demonstrated a graphene plasma resonance capacitor with a resonant frequency  $f_0\simeq 40\;\mathrm{GHz}$ and a quality factor $Q\simeq2$. Quality factors remain smaller than the $Q\sim130$ reported in the ultrashort wavelength ($\lambda_p=0.1$--$0.2\;\mathrm{\mu m}$) MIR domain of Ref.[\onlinecite{Ni2018nature}]. However the GHz damping length ($28\;\mathrm{\mu m}$) is comparable to the MIR value  ($10\;\mathrm{\mu m}$ in Ref.[\onlinecite{Ni2018nature}]), which is promising for applications. We have measured the doping dependence of plasmon velocity, line capacitance and kinetic inductance in good agreement with theory beside a small shift of the kinetic inductance.  Our experiment paves the way to the realization of active plasma resonance transistors working in the $0.1\rightarrow1\;\mathrm{THz}$ domain, above the natural cutoff $\sim 0.1\;\mathrm{THz}$ of conventional graphene field-effect transistors [\onlinecite{Pallecchi2011apl,Mele2018eml}]. Building on this first demonstration performed at cryogenic temperature, a room temperature variant can be envisioned by scaling down the sample size and the plasmon wave length by a factor $10$ to accommodate the  phonon-limited mean free path of $0.7\;\mathrm{\mu m}$ at $300\;\mathrm{K}$. Such an achievement would in particular allow  realizing  plasma resonance transistors in the $600\;\mathrm{GHz}$ frequency domain, highly desirable for high-resolution air-craft RADARs operating in the $mm$-range. The long plasmonic channels are compatible with the incorporation of bottom gate arrays to engineer doping modulation profiles and investigate the propagation of Dirac plasmons in bipolar superlattices.

\section*{Methods}

Micromechanical exfoliation provides a pristine monolayer graphene, which is subsequently encapsulated between two layers of hexagonal boron nitride (hBN) in order to achieve a high electronic mobility (phonon limited at room temperature). The encapsulation is performed by means of a dry pick-up technique using in a polyvinyl alcohol (PVA) and polymethylmethacrylate (PMMA) stamp on a polydimethylsiloxane (PDMS) support [\onlinecite{Banszerus2016nl}]. A first top hBN flake ($10$ to $30\;\mathrm{nm}$ thick) is transferred from its substrate to the PMMA. Then the graphene is picked up by the first hBN flake thanks to the strong van der Waals interactions and deposited on a second bottom hBN flake, which was previously exfoliated onto a high resistivity silicon substrate. The transfer process is carried out using a custom made alignment system with a heating plate operating between $30$ and $130\;^\circ\mathrm{C}$. Finally, the PVA and PMMA polymers are dissolved in hot water ($95\;^\circ\mathrm{C}$) and acetone, respectively.

\begin{figure}[hh]
\centerline{\includegraphics[width=17cm]{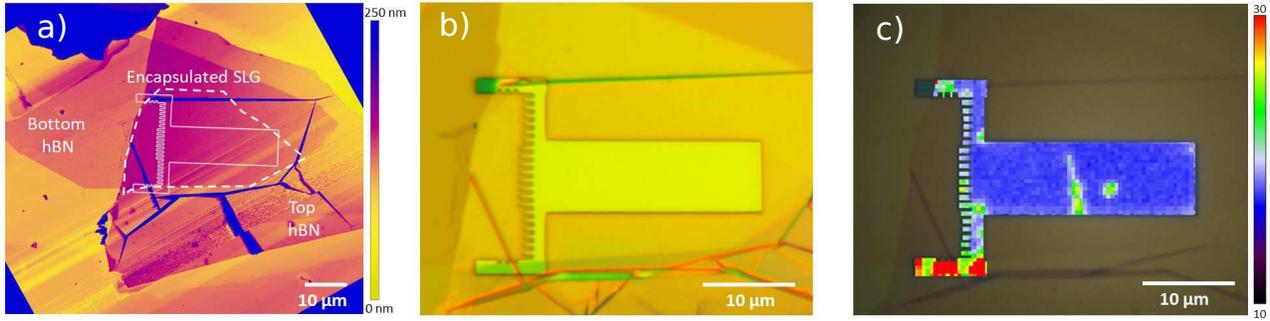}}
\caption{Fabrication process of a plasma resonance capacitor (PRC). a) AFM image of the hBN/graphene/hBN stamp deposited on the high-resistivity  Si0$_2$/Si substrate. The dashed and solid lines show the contours of the pristine graphene flake and targeted PRC sample. b) optical image of the PRC sample after etching. c) Imaging of the Raman 2D-peak width (color encoded in the range $10$--$30\;\mathrm{cm^{-1}}$) showing the good sample homogeneity and highlighting the presence of a spurious fold and bubble in this sample (PRC7).}
\label{PRC_techno.fig3}
\end{figure}

The encapsulated graphene samples are generally hundreds of $\mathrm{\mu m^2}$ in size and need to be patterned into a passive circuit. After characterization by Raman spectroscopy (fig. \ref{PRC_techno.fig3}c) and atomic force microscopy (AFM, fig. \ref{PRC_techno.fig3}a), we used e-beam nanolithography to define a T-shape capacitor (fig. \ref{PRC_techno.fig3}b). The length of the channel, according to its mobility, defines the quarter wave plasma resonance and needs to be long enough to obtain a signal in the $0-40\;\mathrm{GHz}$  bandwidth. The etching of the hBN-graphene-hBN sandwich is performed using a mixture of CHF$_3$/O$_2$ plasma etching through a temporary $40\;\mathrm{nm}$ thick aluminium mask. A chromium/gold ($5$/$200\;\mathrm{nm}$) edge contact is then deposited on the source edge of the T-shape heterostructures (using a comb design to enhance the contact length and reduce the contact resistance). In order to passivate the other edges of the heterostructures and to avoid any source-gate leakage current in the PRC, we oxidized $2\;\mathrm{nm}$ of aluminum followed by $10\;\mathrm{nm}$ of Al$_2$O$_3$ by atomic layer deposition (ALD). Finally, a chromium/gold gate electrode is deposited on the top and the PRC is embedded into a coplanar waveguide (CPW). Similarly, thruline and dummy reference structures were defined on the same chip for de-embedding (see below).

High frequency admittance measurements were carried out in a Janis cryogenic probe station in the temperature range $T=10$--$300\;\mathrm{K}$.  The two-port scattering parameters $S_{ij}$ of the capacitor were measured using an Anritsu MS4644B vector network analyzer (VNA) in the $0-40$ GHz range.  Bias tees were used to decouple the DC gate voltages from the GHz probe signal. A short-open-load-reciprocal (SOLR) protocol was used to calibrate the wave propagation until the probe tips. We then measured $S$ parameters of a symmetric thruline reference structure, calculated its $ABCD$ (cascade) matrix $A_{thru}$ and took the inverse of the square root of this matrix $A_{thru}^{-1/2}$. The wave propagation in the coplanar access of the PRC can now be de-embedded from its $ABCD$ matrix: $A_{PRC}^{deem} = A_{thru}^{-1/2} \cdot A_{PRC} \cdot A_{thru}^{-1/2}$. The same procedure was carried out on a dummy reference structure which has the same contact and gate geometry as the PRC but does not contain encapsulated graphene. Finally, the $ABCD$ matrices of the PRC and the dummy structure were converted to admittance ($Y$) parameters and the $Y$ matrix of the dummy structure was subtracted from that of the PRC in order to de-embed remaining stray capacitances. The $Y_{i,j}$ parameters should now all be the same (except for a minus sign in front of the off-diagonal elements), due to the symmetry of our two-port network. The admittance data shown in this article correspond to one of the off-diagonal elements $-Y_{12}$, which was further de-embedded from the contact resistance: $Y = (-1/Y_{12}-R_c)^{-1}$.

\begin{acknowledgments}
The research leading to these results have received partial funding from the the European Union ``Horizon 2020'' research and innovation programme under grant agreement No. 785219 ``Graphene Core'', and from the ANR-14-CE08-018-05 "GoBN".
\end{acknowledgments}

\newpage

\end{document}